 \documentclass[aps,prl,twocolumn,groupedaddress,amsmath,amssymb,showpacs]{revtex4}
\usepackage{graphicx}
\usepackage{amsmath}
\usepackage{dcolumn}
\usepackage{bm}

\newcommand{\bk}{\mathbf{k}}
\newcommand{\br}{\mathbf{r}}

\newcommand{\ra}{\rangle}
\newcommand{\la}{\langle}

\newcommand{\vx}{\bm x}
\newcommand{\vy}{\bm y}
\newcommand{\vz}{\bm z}

\newcommand{\vs}{\bm s}

\begin{document}
\title{Physical Bounds to the Entropy-Depolarization Relation in Random Light Scattering}
\author{A. Aiello}
\author{J.P. Woerdman}
\affiliation{Huygens Laboratory, Leiden University\\
P.O.\ Box 9504, 2300 RA Leiden, The Netherlands}
\begin{abstract} We present a theoretical study of multi-mode scattering of light  by
optically random  media, using  the Mueller-Stokes formalism which
permits to encode all the polarization properties of the
scattering medium in a real $4 \times 4$ matrix. From this matrix
two relevant parameters can be extracted: the depolarizing power
$D_M$ and the polarization entropy $E_M$ of the scattering medium.
By studying the relation between $E_M$ and $D_M$, we find that
{\em all} scattering media must satisfy some {\em universal}
constraints. These constraints apply to both classical and quantum
scattering processes. The results obtained here may be especially
relevant for quantum communication applications, where
depolarization is synonymous with decoherence.
\end{abstract}
 \maketitle
%
%
%
%
%

%
%
%
\paragraph{Introduction}
The polarization aspects of random light scattering have drawn
quite some interest in recent years, since they present a
diagnostic method of the medium involved and also help
visualization of objects that are hidden inside the medium
\cite{DOP}. When polarized light is incident on an optically
random medium it suffers multiple scattering and, as a result, it
may emerge partly or completely depolarized. The amount of
depolarization can be quantified by calculating
 either the entropy ($E_F$)  or
 the degree of polarization  ($P_F$) of the scattered field \cite{KligerBook}.
It is simple to show that the field quantities  $E_F$ and $P_F$
are related by a single-valued function: $E_F = E_F(P_F)$. For
example polarized light ($P_F = 1$) has $E_F =0$ while partially
polarized light ($0 \leq P_F < 1$) has $1 \geq  E_F > 0$. When the
incident beam is polarized  and the output beam is partially
polarized, the medium is said to be depolarizing. An average
measure of the depolarizing power of the medium is given by the so
called { depolarization index} ($D_M$) \cite{Gil86}.
Non-depolarizing media are characterized by $D_M=1$, while
depolarizing media have $0 \leq D_M<1$. A depolarizing scattering
process is always accompanied by an increase of the entropy of the
light, the increase being due to the interaction of the field with
the medium. An average measure of the entropy that a given random
medium can add to the entropy of the incident light beam, is given
by the {polarization entropy} $E_M$ \cite{LeRoy}. Non-depolarizing
media are characterized by $E_M = 0 $, while for depolarizing
media  $0< E_M \leq 1$. As the field quantities $E_F$ and $P_F$
are related to each other, so are the medium quantities $E_M$ and
$D_M$ with the key difference that, as we shall show later, $E_M$
is a {\em multi}-valued function of $D_M$.

The purpose of this Letter is to point out a universal relation
between the polarization entropy $E_M$ and the depolarization
index $D_M$ valid for any random scattering medium. This relation
covers the complete regime from zero to total depolarization. It
has been introduced before, by Le Roy-Brehonnet and Le Jeune
\cite{LeRoy}, in an empirical sense, to classify depolarization
measurements on rough surfaces (sand, rusty steel, polished steel,
\ldots). We derive here its theoretical foundation and present
analytical expressions for the multi-valued function $E_M =
E_M(D_M)$. Although the $(E_M,D_M)$ relation is essentially
classical, we use a single-photon theoretical approach, exploiting
the well known analogy between single-photon and classical optics
\cite{MandelBook} . We prefer this to a classical formulation
since it offers a natural starting point for extension to
entangled twin-photon light scattering by a random medium, which
{\em is} a true quantum phenomenon that could deteriorate quantum
communication.
\paragraph{Polarization description of the field}
Let us consider a collimated light beam propagating in the
direction $\vz$. In a given spatial point $\br$, the monochromatic
time-dependent electric field associated with the beam is a
complex-valued  vector $\mathbf{E}(t) = X(t)\vx + Y(t) \vy$. This
vector defines the { instantaneous} polarization of the light
which is, in any short enough time interval, fully polarized.
Alternatively, the same light beam may be described by a time
dependent { real-valued} unit Stokes vector $\vs(t) = \{2
\mathrm{Re}(X^* Y), 2 \mathrm{Im} (X^* Y), |X|^2 - |Y|^2 \}/(|X|^2
+ |Y|^2 )$, which moves on the
 Poincar\`{e} sphere (PS) \cite{BornWolf}. Of course,
no detector can  measure the instantaneous polarization, the best
one can get is an average polarization over some time interval
$T$. If during the measurement time $T$ the Stokes vector $\vs(t)$
maintains the same direction, then the beam is polarized. Vice
versa, if $\vs(t)$ moves over the PS covering some finite area,
then
 the beam is partially polarized. In the last case, for stationary
 beams, the motion of $\vs(t)$
produces a probability distribution  over the PS which determines
the degree of polarization of the light \cite{Picozzi04}. Time
dependence of the polarization is not the only cause for
depolarization, also spatial dependence, for example,  may lead to
loss of polarization.

We stress that this picture is not limited to the classical
domain;
 in Ref. \cite{Aiello04_1}  we found, e.g., that a multi-mode
single-photon scattering process generates a $\bk$-dependent
Stokes vector distribution. More generally, if $\psi = \{t, \bk,
\lambda,  \ldots \}$  denotes the set of all variables (e.g., time
$t$, momentum $\bk$, polarization $\lambda$, \dots) on which $\vs
= \vs(\psi)$ depends, then the state of a polarized light beam
(either classical or quantum), may be described by a $2 \times 2$
matrix $\rho(\psi) = (\sigma_0 + \vs(\psi) \cdot \bm{\sigma})/2$,
where $\sigma_0$ is the $2 \times 2$ identity matrix and ${\bm
\sigma} = \{\sigma_1, \sigma_2, \sigma_3 \}$ are the Pauli
matrices. The matrix $\rho(\psi)$ is known as the {\em coherency}
matrix in classical optics \cite{BornWolf} and as the {\em
density} matrix in quantum mechanics \cite{DauFieldRel}. Since by
construction $\mathrm{Tr} \rho(\psi) = 1$, each matrix
$\rho(\psi)$ can describe either a {\em purely} polarized beam in
classical optics, or a {\em pure} photon state in quantum optics
\cite{Fano57}. However, the
 state of a partially polarized beam  must be described by the
 matrix $\rho = \int (d \psi) \, w(\psi) \rho(\psi)$,
 where $\int (d \psi)$ is the integration measure \cite{Nota} in the space of the
variables $\psi$ and $\int (d \psi) \, w(\psi)=1$. The statistical
weight $w(\psi) \geq 0$, defines a probability distribution over
the PS. It is clear that $\rho$ can represent a {\em mixed} photon
state in the context of quantum optics as well. If $A$ denotes any
polarization-dependent observable, its average value must be
calculated as:
\begin{equation}\label{5}
\la {A} \ra = \mathrm{Tr}(\rho A) = \int (d \psi) \, w(\psi)
\mathrm{Tr}({A} \rho(\psi)).
\end{equation}
 If $A$ represents
the entropy of the field, i.e. $A = -\log(\rho)$, then $\la A \ra
= -\mathrm{Tr}(\rho \log \rho) $, which is the von Neumann entropy
$S$ of the photon state \cite{PeresBook}. However, by using Eq.
(\ref{5}) it is easy to see that  this  coincides with the Gibbs
entropy \cite{Wehrl78} of the distribution $w(\psi)$, since $S = -
\int (d \psi) \, w(\psi) \log ( w(\psi))$,  in agreement with the
results of Ref. \cite{Picozzi04}.
\paragraph{Single-photon scattering and multi-mode Mueller formalism}
The theoretical framework for studying one-photon scattering has
been established elsewhere \cite{Aiello04_1}, here we use the
results found in \cite{Aiello04_1} to extend the Mueller-Stokes
formalism to quantum scattering processes. In classical optics a
polarization scattering process can be characterized  by a {
real-valued} $4 \times 4$ matrix, the so called Mueller matrix $M$
\cite{KligerBook}, which describes the polarization properties of
the scattering medium. We show now that such a matrix description
can be extended to the quantum (single-photon) scattering case.
Let us consider a photon prepared in the pure state
${\rho}(\psi)$, approximatively described by a monochromatic plane
wave $|\bk_0, \lambda_0 \ra$. In this case $\psi = \{ \bk_0,
\lambda_0 \}$. Now, let us suppose that the photon is transmitted
through a linear optical system described by an unitary scattering
operator ${\mathcal{T}}$ such that $ {\rho}(\psi') =
{\mathcal{T}}^\dagger {\rho}(\psi) {\mathcal{T}} $ represents the
pure state of the photon after the scattering, where $\psi'$ is
the set of {\em all} scattered modes: $\psi' = \{ \bk_1 ,
\lambda_1 , \bk_2 , \lambda_2, \ldots \}$. A multi-mode detection
scheme implies a reduction from the set $\psi'$  to the subset of
the {\em detected} modes $\psi'' = \{ \bk_1 , \lambda_1 ,\ldots,
\bk_N , \lambda_N \} \subset \psi'$ which causes a transition from
the pure state $\rho(\psi')$ to the mixed state $\rho = \int (d
\psi'') \, w(\psi'') \rho(\psi'')$.
If  we denote the Stokes parameters of the beam before and after
the scattering with $s_\mu = \mathrm{Tr}({\rho}(\psi) \sigma_\mu )
$ and $s_\mu'$  respectively ($\mu = 0,1,2,3$), then the classical
result $s_\mu' = \sum_{\nu = 0}^3 M_{\mu \nu} s_\nu$ is retrieved,
with the difference that a {\em generalized} (measured) Mueller
matrix $||M_{\mu \nu}||$ appears which is defined as
\begin{equation}\label{90}
M_{\mu \nu} \propto \int_\mathcal{\psi''} d \, \bk \, m_{\mu
\nu}(\bk).
\end{equation}
The local (with respect to the momentum) matrix elements $m_{\mu
\nu}(\bk)$ are defined by means of the matrix relation
\begin{equation}\label{100}
W^T (\bk) \mathcal{T} (\bk,\bk_0) \sigma_\mu \mathcal{T}^\dagger
(\bk_0,\bk)W (\bk)=
m_{\mu \nu}(\bk) \sigma_\nu,
\end{equation}
$(\mu,\nu = 0,1,2,3)$, and summation over repeated indices is
understood.  Explicit expressions for the $2 \times 2$ matrices
$W(\bk)$ and $\mathcal{T} (\bk,\bk')$ can be found in Ref.
\cite{Aiello04_1}. The proportionality factor  in Eq. (\ref{90})
can be fixed by imposing the condition $M_{00} = 1$. When
$\mathcal{\psi''}$ reduces to a single mode $\{ \bk, \lambda \}$,
then $ W_{ij}(\bk) = \delta_{ij}$ and the classical formalism is
fully recovered.
\paragraph{Depolarization index $D_M$ and polarization entropy $E_M$}
Now that we have a recipe to calculate the Mueller matrix
describing a multi-mode scattering process, we use this knowledge
to study the depolarization properties of the scattering medium.
Within the Mueller-Stokes formalism, the degree of polarization
$P_F$ of the field and the depolarization index $D_M$ of the
medium, are defined as $P_F = (s_1^2 + s_2^2 +s_3^2)^{1/2}/s_0$
and $D_M = \left( \mathrm{Tr}(M^T M)/3 - 1/3 \right)^{1/2}$,
respectively, where $s_\mu$ ($\mu = 0,1,2,3$) are the  Stokes
parameters of the field and $M_{00}=1$ has been assumed.
 A deeper characterization of the scattering medium can be
achieved by using the Hermitian matrix $H$
\cite{Simon82,Anderson94} defined as
\begin{equation}\label{130}
H = \frac{1}{4} \sum_{\mu,\nu}^{0,3} M_{\mu \nu} \left( \sigma_\mu
\otimes \sigma_\nu^* \right),
\end{equation}
where $\mathrm{Tr}(H) =1$. It can be shown \cite{LeRoy} that a
physically realizable optical system is characterized by a
positive-semidefinite  matrix $H$. Let $0 \leq \lambda_\nu \leq 1$
($\nu = 0,\ldots,3$) be the  eigenvalues of $H$. Then it is
possible to express both the depolarization index $D_M$ and the
polarization entropy $E_M$ as a function of the $\lambda_\nu$'s.
Explicitly we have
\begin{equation}\label{140}
D_M = \left[ \frac{1}{3}\left( 4 \sum_{\nu=0}^3 \lambda_\nu^2 - 1
\right) \right]^{1/2},
\end{equation}
and
\begin{equation}\label{150}
E_M = -\sum_{\nu=0}^3 \lambda_\nu \log_4(\lambda_\nu ).
\end{equation}

Now we are ready to show the  universal character of the $(E_M,
D_M)$ plot originally introduced in Ref. \cite{LeRoy}. More
precisely, we show that it allows to characterize {\em all}
possible scattering media by means of their polarimetric
properties. The main idea is the following: both $E_M$ and $D_M$
depend on the four real eigenvalues of $H$ which actually reduces
to three independent variables because of the trace constraint
$\mathrm{Tr}(H) =1$. If we use Eq. (\ref{140}) to eliminate one of
these variables in favour of $D_M$ we can write $E_M =
E_M(D_M,\alpha, \beta)$ where $\alpha,\beta$ represent the last
two independent variables. Then, for each value of $0 \leq D_M
\leq 1$, different values of $E_M$ can be obtained by varying
$\alpha$ and $\beta$ between $0$ and $1$. In such a way we obtain
a whole domain in the $D_M$-$E_M$ plane instead of just a curve.
In order to do that, we have implemented a Monte Carlo code to
generate a uniform distribution of points over the 4-dimensional
unit sphere: the square of the four coordinates of each point is
an admissible set of eigenvalues of $H$. In this way we have
generated the graph shown in Fig. 1.
\begin{figure}[!ht]
\includegraphics[angle=0,width=7truecm]{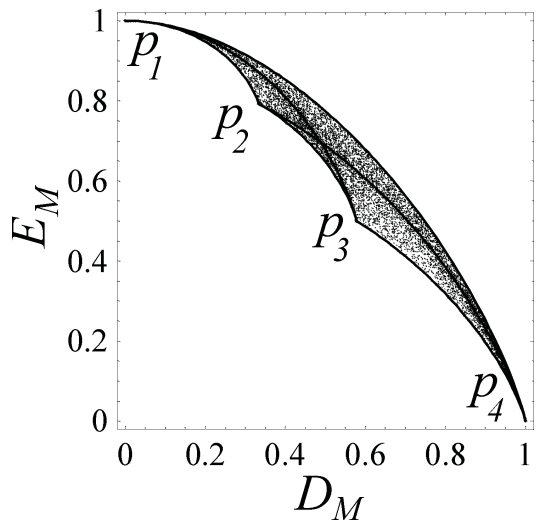}
\caption{\label{fig:1}  Numerically determined domain in the $D_M
-E_M$ plane corresponding to all physically realizable
polarization scattering processes. The solid curves are the
analytically obtained bounds. The four cusp points $p_1 = (0,1)$,
$p_2 = (1/3, \log_4 3 )$, $p_3 = (1/\sqrt{3},1/2)$, $p_4 = (1,0)$
separate different polarization scattering processes, as described
in the text.}
\end{figure}
The boundary of this domain is formed by the curves
$\mathcal{C}_{ij}$ ($i,j = 1,\ldots,4$), joining the points $(p_i
\rightarrow p_j)$. The analytical expressions for these curves are
\begin{equation}\label{180}
E(n,f) =  -\left[ \left(1 - n f \right) \log_4 \left(1 - n f
\right) + \, n f \log_4 \left( f \right) \right],
\end{equation}
where
\begin{equation}\label{190}
f_\pm = \frac{1}{n+1}\left[ 1 \pm \sqrt{1 - \frac{3}{4}
\frac{n+1}{n} (1 - D_M^2)} \right].
\end{equation}
The links between the functions $E(n,f)$ and the curves
$\mathcal{C}_{ij}$ are given in Table I where we have defined
 $E_{13} = -(1-\mu) \log_4(\frac{1-\mu}{2}) - \mu
\log_4(\frac{\mu}{2}) $.
\begin{table}
\caption{\label{ta_1} List of the analytical curves (continuous
lines) in Fig. 1. The second column refers to the equations
generating the corresponding curves, while the third column gives
the eigenvalues of $H$. The first four curves form the boundary of
the physical domain; the last two represent inside curves.}
\begin{tabular}{c c c c c}
  \hline \hline
    Curve & & Generating equation  & & Eigenvalues of $H$  \\
     \hline
  $\mathcal{C}_{12}$ & \phantom{.....} & $E(3,f_+)$ & \phantom{.....} &
  $\{\lambda, \mu, \mu,\mu \}$   \\
  $\mathcal{C}_{23}$ & & $E(2,f_+)$ &  & $\{\lambda, \mu, \mu,0 \}$   \\
  $\mathcal{C}_{34}$ & & $E(1,f_\pm)$ & & $\{\lambda, \mu, 0,0 \}$  \\
    $\mathcal{C}_{14}$ & & $E(3,f_-)$ & & $\{\lambda, \mu, \mu,\mu \}$
    \\
\hline
    $\mathcal{C}_{13}$ & & $E_{13}$ & & $\{\lambda, \lambda, \mu,\mu \}$   \\
  $\mathcal{C}_{24}$ & & $E(2,f_-)$ & & $\{\lambda, \mu, \mu,0 \}$   \\
  \hline \hline
\end{tabular}
\end{table}
The curve $\mathcal{C}_{14}$ is special in the sense that it sets
an upper bound for the entropy of {\em any} scattering medium. We
find numerically that the value of the entropy on this curve is
very well approximated by
\begin{equation}\label{170}
E_M^\mathrm{cr} \sim \left( 1 -D_M^2 \right)^\gamma,
\end{equation}
where $\gamma \cong 0.862$, which is, interestingly, almost equal
to $e/\pi$. Then, for all depolarizing scattering media the
condition $E_M \lesssim E_M^\mathrm{cr}$ must be satisfied.
It is interesting to note that a purely depolarizing scattering
medium (with diagonal Mueller matrix) leads to $E_M \cong E_C$. By
using  thermodynamics language, one may interpret  Fig. 1
 as a polarization ``state diagram'' where
different phases of a generic scattering medium, characterized by
different symmetries of the corresponding Mueller matrices, are
separated by the curves $\mathcal{C}_{ij}$. It is worth to note
again that there is nothing inherently quantum in the above
derivation of the physical bounds Eq. (\ref{180}), therefore these
results have validity both in the classical and in the quantum
regime.
\paragraph{Random matrix approach}
We have checked the validity of the theory outlined above, for
scattering media in the regime of applicability of the
random-matrix theory (RMT) \cite{MehtaBook}. Random media, either
disordered media \cite{DOP} or chaotic optical cavities
\cite{Aiello03}, can be represented by ensembles \cite{Kim87}. The
transmission of polarized light through a random medium may
decrease the degree of polarization in a way that depends on the
number $N$ of the detected modes via Eq. (\ref{90}). Under certain
conditions, RMT can account for a statistical description of the
light scattering by random media \cite{Beenakker97_1,Guhr98}. Let
$\psi_\lambda(\bk)$ be the complex probability amplitude that a
photon is scattered in the state $|\bk, \lambda \ra$. Then,
according to RMT, the real and the imaginary parts of the
scattering amplitudes $\psi_\lambda(\bk)$ are independent Gaussian
random variables with zero mean and variance that can be fixed to
$1$. The assumption of independent variables is justified since
usually the set $\psi''$ of the detected modes is much smaller
than the set $\psi'$ of the {\em all} scattered modes
\cite{Velsen04}. Let us suppose now that the impinging photon is
in the pure state $| \bk_0, \lambda_0 \ra$. In this case
$\psi_\lambda(\bk) = \mathcal{T}_{\lambda \lambda_0}(\bk, \bk_0)$
and the statistical distribution of the $M_{\mu \nu}$'s can be
numerically calculated accordingly to Eqs. (\ref{90}-\ref{100}).
In this way we have calculated the ensemble-averaged polarization
entropy $\la E_M \ra$ and depolarization index $\la D_M \ra$ of
the medium, as functions of  $N$ for the case in which the angular
aperture of the detector is so small that $W_{ij}(\bk) \simeq
\delta_{ij}$. The results are shown in Fig. 2 for the cases of  a
generic scattering medium ($\mathcal{T}_{\lambda \lambda_0}(\bk,
\bk_0)$ unconstrained) and of a polarization-conserving medium
($\mathcal{T}_{\lambda \lambda_0}(\bk, \bk_0) \propto
\delta_{\lambda \lambda_0}$). The last case is realized when the
geometry of the scattering process is confined  in a
 plane. As one can see, for both cases RMT results cover only a small part
of the ($E_M,D_M$) diagram; however, the numerical data are
consistent with the analytical bounds given by Eq. (\ref{180}).
\begin{figure}[!ht]
\includegraphics[angle=0,width=7truecm]{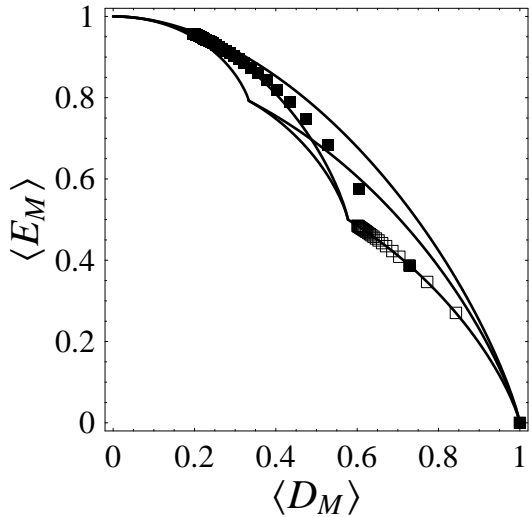}
\caption{\label{fig:2} RMT results for the ensemble-averaged
polarization entropy $\la E_M \ra$ as function of the
ensemble-averaged depolarization-index $\la D_M \ra$ for generic
(dark squares) and polarization-conserving (open squares)
scattering processes. In both cases each point correspond to a
given number $N$ of detected modes. When $N$ increases from $1$ to
$30$, points move from the bottom to the top of the figure. The
solid lines are the analytical bounds of Fig. 1.}
\end{figure}
\paragraph{Conclusions}
In summary, we have studied  the scattering of light by optically
random media, from a polarization point of view. To this end we
have first extended the Mueller-Stokes formalism to make it
suitable for the description of single-photon scattering
processes. Then, after the calculation of the Mueller matrix $M$
characterizing the polarization properties of the scattering
medium, we have extracted from $M$ the depolarization index $D_M$
and the polarization entropy $E_M$. By analyzing the functional
relation between $E_M$ and $D_M$, we have found that the
polarization properties of any scattering medium are constrained
by some physical bounds. These bounds have an universal character
and they hold in both the classical and the quantum regime. Our
results provide a deeper insight in the nature of random light
scattering by giving an useful tool, both to theoreticians and
experimenters, to classify scattering media according to their
polarization properties. The use of this tool may  be particularly
relevant in
 quantum communication where it is desirable to
manipulate and control the polarization of the light
\cite{Legre03}. Presently, experiments are in progress in our
group to verify this theoretical framework.
\begin{acknowledgments}
 We acknowledge support from the EU under the
IST-ATESIT contract. This project is also supported by FOM.
\end{acknowledgments}


\end{document}